\documentclass[runningheads]{llncs}
\usepackage{graphicx}
\usepackage{verbatim}
\usepackage{listings}
\usepackage{xcolor}
\usepackage{wrapfig}
\PassOptionsToPackage{hyphens}{url}\usepackage{hyperref}
\usepackage{float}
\usepackage{caption}
\usepackage{subcaption}

\definecolor{mGray}{rgb}{0.5,0.5,0.5}
\def \PosDBAnonymized {PosDB}

\begin{document}
\title{Finding a Second Wind: Speeding Up Graph Traversal Queries in RDBMSs Using Column-Oriented Processing}
%
%
\author{Mikhail Firsov\inst{1}\orcidID{0000-0001-6739-7303} \and
Michael Polyntsov\inst{1}\orcidID{0000-0001-7356-2504} \and Kirill Smirnov\inst{1}\orcidID{0000-0003-4727-3455} \and
George Chernishev\inst{1}\orcidID{0000-0002-4265-9642}}
\authorrunning{M. Firsov et al.}
%
\institute{
Saint-Petersburg University, Russia\\
\email{\{mikhail.a.firsov, polyntsov.m, kirill.k.smirnov, chernishev\}@gmail.com}
}
\maketitle              

\begin{abstract}

Recursive queries and recursive derived tables constitute an important part of the SQL standard. Their efficient processing is important for many real-life applications that rely on graph or hierarchy traversal. Position-enabled column-stores offer a novel opportunity to improve run times for this type of queries. Such systems allow the engine to explicitly use data positions (row ids) inside its core and thus, enable novel efficient implementations of query plan operators.

In this paper, we present an approach that significantly speeds up recursive query processing inside RDBMSes. Its core idea is to employ a particular aspect of column-store technology (late materialization) which enables the query engine to manipulate data positions during query execution. Based on it, we propose two sets of Volcano-style operators intended to process different query cases.

In order validate our ideas, we have implemented the proposed approach in \PosDBAnonymized{}, an RDBMS column-store with SQL support. We experimentally demonstrate the viability of our approach by providing a comparison with PostgreSQL. Experiments show that for breadth-first search: 1) our position-based approach yields up to 6x better results than PostgreSQL, 2) our tuple-based one results in only 3x improvement when using a special rewriting technique, but it can work in a larger number of cases, and 3) both approaches can't be emulated in row-stores efficiently.

\keywords{Query Processing \and Column-stores \and Recursive Queries \and Late Materialization \and Breadth-First Search \and PosDB.}
\end{abstract}

%
%
%

\section{Introduction}

The ANSI’99 SQL standard introduced the concept of recursion into SQL with syntactic constructs to define recursive views and recursive derived tables. This allows users to store graph data in a tabular form and to express some graph queries using CTEs and recursive syntax. The admissible subset is rather limited compared to the specialized graph systems, but it is sufficient to solve a number of common tasks. Such tasks originate from many real-life applications and usually concern some hierarchy traversal which comes as a breadth-first search computation.

In this paper, we present another outlook on RDBMS architecture that significantly improves system performance at least for some types of graph queries expressed by recursive SQL. More specifically, we present a column-oriented approach that will improve run times for queries that perform breadth-first search.

Having emerged about fifteen years ago, column-stores quickly became ubiquitous in analytic processing of relational data. Their idea is simple: store data in a columnar form in order to read only necessary columns for evaluating the query. Such approach also provides better data compression rates~\cite{Abadi:2013:DIM:2602024}, improves CPU cache utilization, facilitates SIMD-enabled data processing, and offers other benefits. However, some column-stores additionally allow the query engine to explicitly use data positions during query execution. This made way for a number of optimizations and techniques that offered various benefits for query processing. Thus, we differentiate the ``position-enabled'' column-stores from the rest as the column-stores that are able to reap benefits from explicit position manipulation inside their engine. We believe that such an approach can give RDBMs a second wind in handling graph queries.

Positions (also called row ids or offsets) are integers that refer to some record or individual attribute value inside a table. Operating on data positions allows query engine to achieve savings by deferring switching to data values. This group of techniques is called late materialization and it was successfully employed for various query plan operators~\cite{c-store,Boncz:1999:MPQ:765509.765511,FlashJoin,Mukhaleva:2019:IWF}.

We employ this technique to design two sets of Volcano-style~\cite{Graefe:1993:QET:152610.152611} operators intended to handle different query cases that involve recursive processing. We have implemented them inside a position-enabled column-store \PosDBAnonymized{}~\cite{DOLAP22-newarch,Chernishev:2018:PAO}. Next, in order to evaluate them we run experiments with queries that perform breadth first search. We have also performed the comparison of our approach with PostgreSQL. 

The overall contribution of this paper is the following:
\begin{enumerate}
    \item A survey of existing query processing techniques in RDBMSs that concern recursive queries.
    \item A design of two query operators for position-enabled column store that speed up recursive query evaluation.
    \item An experimental evaluation of proposed techniques and a comparison with state-of-the-art row-store RDBMS.
\end{enumerate}

This paper is organized as follows. In Section~\ref{sec:relwork} we survey various aspects of implementation and usage of recursive queries inside relational DBMSs. Then, in Section~\ref{sec:background} we present the main features of \PosDBAnonymized{} and discuss its query processing internals. After this, in Section~\ref{sec:solution} we describe implementation details of the proposed recursive operators and their use in the existing query plan model of \PosDBAnonymized{}. Section~\ref{sec:experiments} contains an  evaluation that compares \PosDBAnonymized{} with PostgreSQL using a series of experiments on trees of different size, height and additional payload. Finally, in Section~\ref{sec:concl} we conclude this paper and discuss future work.

\section{Related Work and Motivation}\label{sec:relwork}

\subsection{Related Work}

In this section we review existing papers that address graph query processing in SQL-supporting systems, paying special attention to recursive evaluation.

One of the earliest papers that addressed the problem of recursive query evaluation was the paper~\cite{graphQueriesOptimize}. There, author introduces several query optimizations for recursive queries with graphs, namely: early evaluation of row selection conditions, elimination of duplicate rows in intermediate tables, and using an index to accelerate join computation.

The authors of the papers~\cite{AdaptiveOptimize,10.1007/11827405-34} describe several issues with query optimization in relational databases when implementing recursive queries. The first approach they mention is the full feedback approach (FFB), which provides the optimizer with the demographics of each recursion iteration so that it can generate a new plan for the subsequent iteration. However, FFB interrupts potential pipelining and cannot take advantage of global query optimizations, making it unsuitable for parallel DBMS. The next approach, look ahead with feedback (LAWF), generates plans for the subsequent $k$ iterations in advance, with k depending on the query planning cost and propagation of join estimation errors. The authors present a dynamic feedback mechanism based on passive monitoring to collect feedback and to determine when re-planning is necessary. The LAWF method supports both pipelining and global query optimization.

In the paper~\cite{largeEvaluate} the authors consider two graph problems: transitive closure computation and adjacency matrix multiplication. In order to solve them, the authors study the optimization of queries that involve recursive joins and recursive aggregations in column- and row-oriented DBMS. They evaluate the impact of several query optimization techniques, such as pushing aggregation through recursion and using ORDER BY with merge joins in column-store instead of hash joins. The authors evaluate effects of indexing columns containing vertices and effects of sorting rows in a row-store to evaluate the iteration of $k$ joins.

In the paper~\cite{linearOptimize} the author evaluates various recursive query optimizations for the plan generator. The paper considers five techniques: storage and indexing for efficient join computation, early selection, early evaluation of non-recursive joins, pushing duplicate row elimination, and pushing aggregation. The author uses four types of graphs: tree, list, cyclic, and complete graphs. However, similarly to previous work~\cite{largeEvaluate} author uses a sequence of SQL commands (including INSERTs) to implement the proposed optimizations. Such approach may suffer from various overheads, as opposed to implementing an operator node in the engine source code. This, in turn, may lead to inaccurate results.

The Recursive-aggregate-SQL (RaSQL)~\cite{RaSQL} system extends the current SQL standard to support aggregates in recursion. It can express powerful queries and declarative algorithms, such as graph and data mining algorithms. The RaSQL compiler allows mapping declarative queries into one basic fixpoint operator supporting aggregates in recursive queries. The aggregate-in-recursion optimization brought by the PreM property and other improvements make the RaSQL system more performant than other similar systems.

In the paper~\cite{hugeProperty} the authors address the problem of storing large property data graphs inside relational DBMS. They 
adapt the SQLGraph~\cite{10.1145/2723372.2723732} approach to reduce the disk volume and increase processing speeds. They evaluate their schema using the PostgreSQL on LDBC-SNB and show that their schema not only performs better on read-only queries but also performs better on workloads that include update operations.

Graph databases are good for storing and querying provenance data. One of the earliest papers that
evaluated this possibility was the study~\cite{traversalOfRelational}. The authors compare relational and graph databases on different types of queries. This study demonstrated that for traversal queries, graph databases were clearly faster, sometimes by a factor of 10. This result was expected since relational databases are not designed to perform traversals such as standard breadth-first-search.

Another paper that concerned graph databases in data provenance domain was the study~\cite{graphVSRelationalDB}. The authors propose an improved version of the DPHQ framework for capturing and querying the provenance data. They conclude that graph databases offer significant performance gains over relational databases for executing multi-depth queries on provenance. The performance gains become much more pronounced with the increase in traversal depth and data volumes.

\subsection{Motivation}

The related works discussed above demonstrated popularity and relevance of graph queries and graph database systems. However, they also showed that there is only a handful of studies that address processing of graph queries (BFS, transitive closure) using the recursion technique in SQL-supporting systems.

Moreover, despite the existence of studies which touch upon the processing of recursive SQL queries in column-stores (e.g.~\cite{largeEvaluate}), there are no studies that propose to leverage data positions. On the other hand, in our paper we propose an in-depth operator redesign, which is based on this idea.

\section{Background}\label{sec:background}

\begin{figure*}[t]
\centering
\begin{subfigure}{.29\textwidth}
    \centering
    \includegraphics[width=0.75\textwidth]{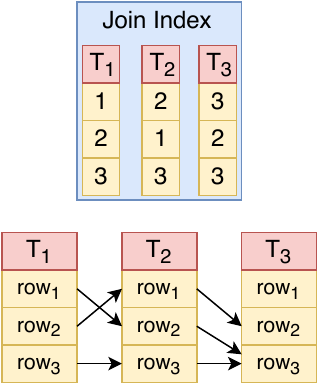}
    \caption{Example of join index}
    \label{fig:join-index}
\end{subfigure}
\begin{subfigure}{.7\textwidth}
    \centering
    \includegraphics[width=0.9\textwidth]{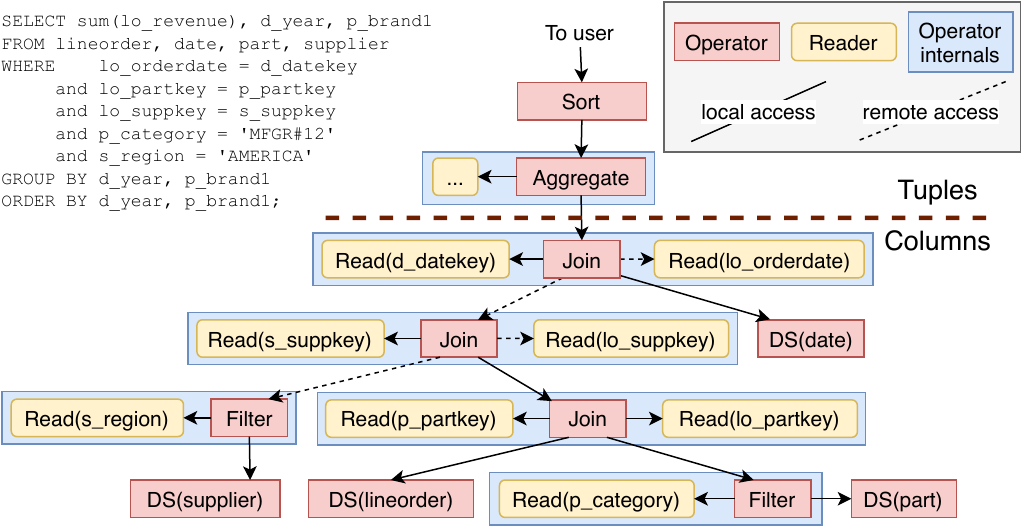}
    \caption{Query plan example}
    \label{fig:plan}
\end{subfigure}
\caption{\PosDBAnonymized{} internals}
\label{fig:test}
\end{figure*}

\PosDBAnonymized{} is a disk-based distributed column-store which features explicit position manipulation, i.e. it is a ``position-enabled'' system. In this regard it is close to the ideas of early systems such as the C-Store~\cite{c-store} and the MonetDB~\cite{monet_2012}. 

\PosDBAnonymized{} uses the pull-based Volcano model~\cite{Graefe:1993:QET:152610.152611} with block-oriented processing. Its core idea is to employ two types of intermediate representations: tuple- and position-based. In the tuple-based representation operators exchange blocks of value tuples. This type of representation is similar to most existing DBMSs. On the other hand, position-based representation is a characteristic feature of \PosDBAnonymized{}. In the positional form, intermediates are represented by a generalized join index~\cite{Valduriez:1987:JI:22952.22955} which is presented in Fig.~\ref{fig:join-index}. Join index stores an array of record indices, i.e. positions, for each table it covers (top of Fig.~\ref{fig:join-index}). Tuples are encoded using rows in the join index. 
Most operators in \PosDBAnonymized{} are either positional or tuple-based, with positional ones having specialized Reader entities for reading values of individual table attributes. The query plan in \PosDBAnonymized{} is divided into positional and tuple parts, and the moment of converting positions into tuples is called materialization. Materialization is to be performed at some moment of query plan, since user needs not positions, but tuples. It can be performed by either a special \texttt{Materialize} operator or by some operators, such as an aggregation operator.

In the query plan presented in Fig.~\ref{fig:plan}, the materialization point is indicated by a brown dotted line. Below the line, positional representation is used and above the line uses tuple representaion. In the latest version of \PosDBAnonymized{}, a query plan may contain several materialization points, in such a manner that every leaf-root path will have one.

Such architecture leads to several different classes of query plans which are discussed in reference~\cite{DOLAP22-newarch}. Operating on positions instead of tuple values allows to achieve significant cost savings for some queries. For example, in case of filtering join it is possible to reduce the total amount of data read from disk if join will be performed on positions first, and then the rest of necessary columns will be read. This is a general idea of late materialization and it was extensively used for implementing many~\cite{Abadi:2013:DIM:2602024,columns_tutorial,Mukhaleva:2019:IWF,TuchinaEtAl:SEIM2018,10.1007/978-3-031-15743-1-46} operators and their combinations. At the same time, in \PosDBAnonymized{} it is possible to build plans equivalent to naive~\cite{Abadi:2013:DIM:2602024} column-stores, i.e. which will read only necessary columns, construct tuples and continue as it was row-store. In this paper we are going to discuss an application of this technique for processing recursive queries.

\PosDBAnonymized{} is a large project and it has many features and implementation details. However, they are out of the scope of this work and are not necessary for its understanding. A detailed description of baseline architecture can be found in paper~\cite{Chernishev:2018:PAO}, and the recent additions are described in~\cite{DOLAP22-newarch}. Finally, an interactive demo of \PosDBAnonymized{} can be seen at the following link\footnote{https://pos-db.com/}.

\section{Proposed Approach}\label{sec:solution}

In order to implement recursive queries in the \PosDBAnonymized{}, we have introduced two new operator groups into its operator set. These groups share the same use pattern and differ only in the used data representation (rows or positions). Their generalized operation flow is presented in Fig.~\ref{fig:realisation}, which is as follows:
\begin{figure}[h!]
    \centering
    \includegraphics[width=0.4\linewidth]{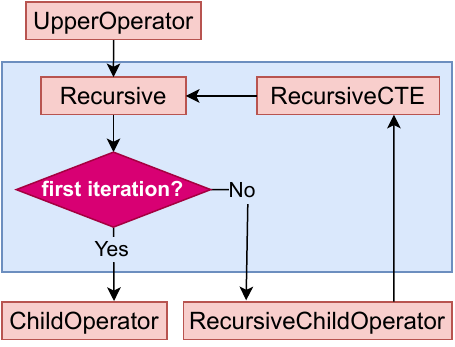}
    \caption{Sample query plan representation using Recursive
and RecursiveCTE}
    \label{fig:realisation}
\end{figure}
\begin{itemize}
    \item The \texttt{Recursive} operator stores pointers to \texttt{RecursiveCTE}, \texttt{ChildOpe\-ra\-tor} and \texttt{RecursiveChildOperator}. \texttt{ChildOperator} is used for the non-recursive part of the query, with its help \PosDBAnonymized{} gets initial rows or initial positions. The \texttt{Re\-cur\-si\-ve\-Child\-Ope\-rator} is a regular operator, but internally it either explicitly or implicitly (via several intermediate operators) receives data from \texttt{RecursiveCTE}.
    \item \texttt{RecursiveCTE} stores a pointer to the Recursive, from which it asks for new records to be passed by the Next method in \texttt{RecursiveChildOperator}.
\end{itemize}

Recall that there are two types of intermediate data representation in \PosDBAnonymized{}: tuple-based and positional. This results in two sets of operators:

\begin{itemize}
     \item \texttt{TRecursive} and \texttt{TRecursiveCTE} that only work with blocks of tuples.
     \item \texttt{PRecursive} and \texttt{PRecursiveCTE} that only work with position blocks.
\end{itemize}

We have designed only these two sets, each focusing on one particular data representation, either tuple-based or positional. However, the first thing which comes to mind is to use a combination of tuple-based and positional operators. For example, consider a case when \texttt{ChildOperator} and \texttt{RecursiveCTE} return a position block and \texttt{RecursiveChildOperator} returns a tuple block. In this case, the query engine will have to translate the tuples received from \texttt{RecursiveChildOperator} back to positions in order to use them for the second and subsequent steps of the recursion. However, this may be impossible in certain circumstances, if, for example, a generated attribute (e.g. $value*2 + 1$) is present in the tuple block. In this case, there will be no original column which may be pointed to by a position.

For the sake of clarity, we are going to work through an example. Consider the following recursive query to find all neighbors of a vertex with id = 0 up to depth 4:
\begin{lstlisting}[language=sql, numberstyle=\color{mGray}, label={lst:query2}, numbers=left, basicstyle=\footnotesize, xleftmargin=2em,frame=single,framexleftmargin=1.5em, caption=Recursive query example]
WITH RECURSIVE edges_cte (id, from, to) AS 
    (SELECT edges.id, edges.from, edges.to
     FROM edges WHERE edges.from = 0
     UNION ALL
     SELECT edges.id, edges.from, edges.to
     FROM edges JOIN  edges_cte AS e
     ON edges.from = e.to_v)
SELECT edges_cte.id, edges_cte.from, edges_cte.to
FROM edges_cte
OPTION (MAXRECURSION 4);
\end{lstlisting}

\begin{figure}[h!]
    \centering
    \includegraphics[width=\linewidth]{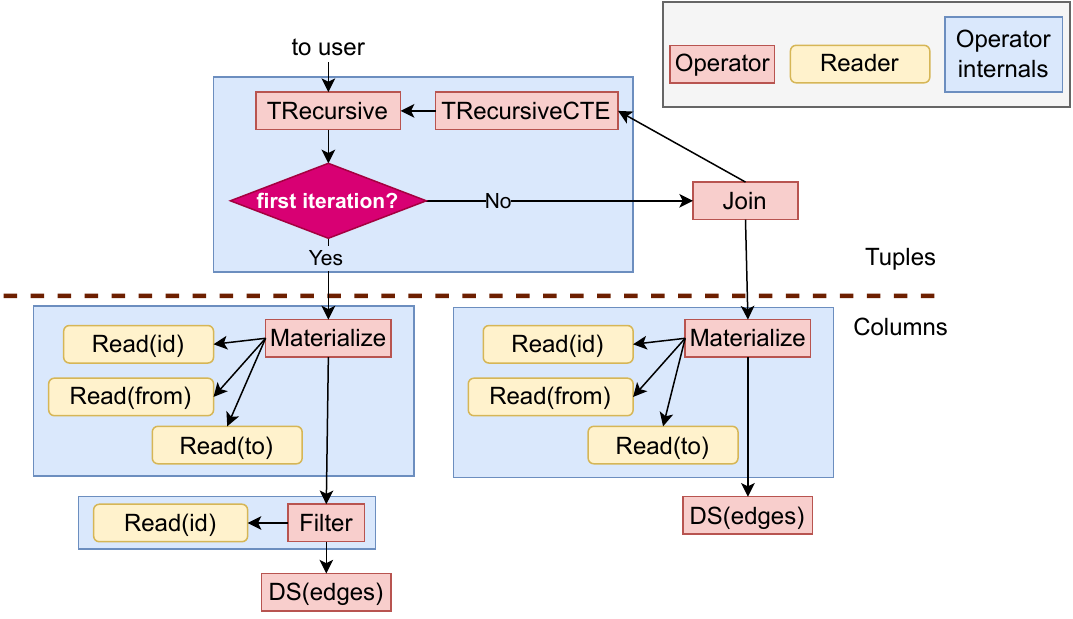}
     \caption{Query tree using TRecursive and TRecursiveCTE}
     \label{fig:exTRecursive}
\end{figure}

The plan of this query using the introduced structures can be represented via the diagram in Fig.~\ref{fig:exTRecursive}. In this figure:

\begin{itemize}
     \item The left \texttt{Materialize} operator is a \texttt{ChildOperator}: it will be executed once in order to initialize the starting set of tuples.
     \item The Join is a \texttt{RecursiveChildOperator}.
     \item The set of tuple blocks of the current recursion step is stored inside \texttt{TRecursive}: we will call it \texttt{curLevel}. In addition, \texttt{TRecursive} will store the position of the block in \texttt{curLevel}, which should be passed to \texttt{TRecursive} next time.
\end{itemize}

The evaluation itself is as follows:

\begin{enumerate}
     \item \texttt{TRecursive} requests blocks from the left \texttt{Materialize} as long as they are not empty and stores them in \texttt{curLevel}.
     \item \texttt{TRecursive} passes all blocks from \texttt{curLevel} up until it reaches the end of \texttt{curLevel}.
     \item To get the block of the new recursion step, \texttt{TRecursive} requests the block from \texttt{Join}.
     \item \texttt{Join} requests blocks from \texttt{TRecursiveCTE} and from the right \texttt{Materialize}.
     \item \texttt{TRecursiveCTE} asks \texttt{TRecursive} for blocks.
     \item \texttt{TRecursive} increments the internal counter and passes \newline \texttt{TRecursiveCTE} in response to its requests for blocks from \texttt{curLevel}.
     \item After typing a new block of a certain size, \texttt{Join} gives it to \texttt{TRecursive}. If it is a non-empty block, then \texttt{TRecursive} will store it in \texttt{nextLevel} temporary storage and proceed to Step 3. If it is an empty block, then \texttt{curLevel} is replaced with \texttt{nextLevel}. If now \texttt{curLevel} is an empty set of blocks, then we say that \texttt{TRecursive} has finished its work, otherwise \texttt{TRecursive} proceeds to Step 2.
\end{enumerate}
\begin{figure}[h!]
     \centering
     \includegraphics[width=\linewidth]{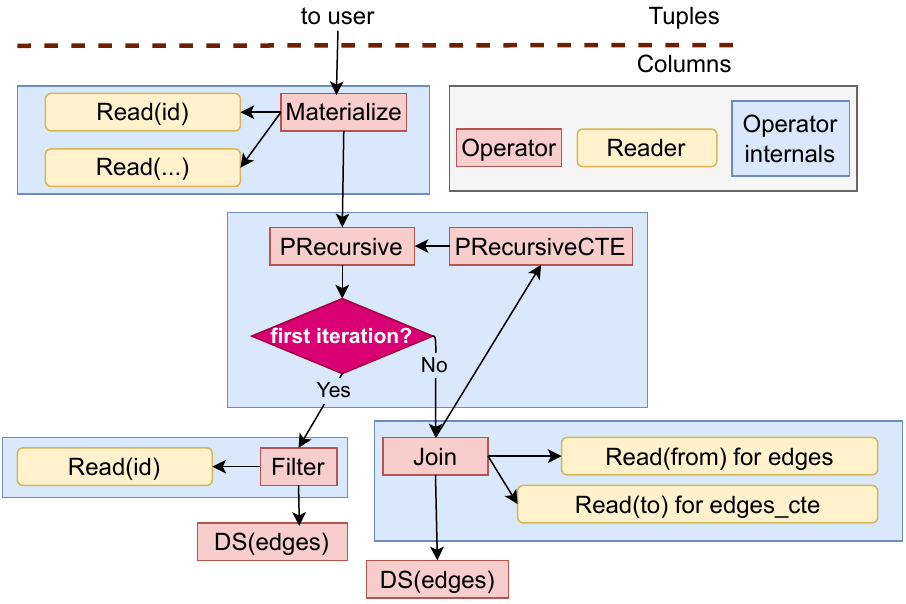}
     \caption{Query tree using PRecursive and PRecursiveCTE}
     \label{fig:exPRecursive}
\end{figure}

The plan of the same query using \texttt{PRecursive} operator can be represented via the diagram in Fig.~\ref{fig:exPRecursive}. Evaluation of this query tree will proceed in a similar way. An important limitation here is that we can only work with positions of the same table, which means that the \texttt{Join} operator must return the positions of the same table as the left \texttt{Filter} operator. However, \texttt{curLevel} stores not tuple blocks, but positions blocks. In all other respects, the logic of the \texttt{PRecursive} and \texttt{PRecursiveCTE} operators is completely identical to the logic of their tuple-based counterparts.

\section{Evaluation}\label{sec:experiments}

\subsection{Methodology}
We evaluate our implementations using hierarchical recursive queries on a tree graph. We generated the datasets for the experiments with a simple script\footnote{\url{https://github.com/Firsov62121/tree_generator}}. All evaluated queries solve the task of finding all nodes that lie at a distance of $n$ hops from the root using the BFS algorithm. A test graph was stored in \PosDBAnonymized{} and PostgreSQL as an edge list. Columns are of the following types: \texttt{id}, \texttt{from}, \texttt{to} are \texttt{int} (4 bytes); \texttt{name} is \texttt{varchar(15)} (32 bytes); each additional column in the second and third test sets is \texttt{varchar(20)} (42 bytes). The number of table rows is indicated above the figures with test results.

In order to evaluate our solution, we have selected a baseline of PostgreSQL. Our choice is based on the following considerations. First of all, we needed a classic row-store system in order to demonstrate the advantages of our approach. Second, PostgreSQL meets another important requirement: it is free from the DeWitt clause\footnote{\url{https://www.brentozar.com/archive/2018/05/the-dewitt-clause-why-you-rarely-see-database-benchmarks/}}. 

PostgreSQL was configured as follows: JIT compilation and parallelism were disabled since JIT compilation is not implemented in \PosDBAnonymized{} and enabling parallelism would add unnecessary complexity without contributing anything important in the scope of this paper. To ensure that hash join is used in the query plan as in \PosDBAnonymized{}, merge and nested loop joins were turned off using planner parameters. The \texttt{temp\_buffers} and \texttt{work\_mem} parameters were set to values that ensure that any table under test fits into memory. To prevent caching from affecting the results, PostgreSQL internal caches were cleared between runs. 

\PosDBAnonymized{} buffer manager was set to 1GB (32K pages of 32KB size).

Each experiment was repeated 10 times, and the average of the results was calculated. 


We pose the following research questions:
\begin{enumerate}
    \item[RQ1] Does our position-based approach bring any performance gain in a special case when all attributes of a table are required in the recursive part of a query?
    \item[RQ2] What performance advantage does our position-based approach offer when introducing additional payload through auxiliary attributes used in projection?
    \item[RQ3] Is it possible to emulate our approach inside a row-store by rewriting a query in such a way that it will keep a minimum number of columns inside the recursive core and then join the rest?
\end{enumerate}

To answer these questions, we have designed the following experiments, corresponding to each RQ:

\begin{enumerate}
    \item For the first experiment, we used a BFS query with the table consisting only of attributes required for the traversal, giving no benefits to \PosDBAnonymized{} (see Listing~\ref{lst:query2} and corresponding plans in Figures~\ref{fig:exTRecursive},\ref{fig:exPRecursive}).
    \item For the second experiment, we used the query from the first experiment modified by adding payload attributes to the input table itself and to all projections. SQL queries of the following type were used for PostgreSQL and \PosDBAnonymized{}:

\begin{lstlisting}[language=sql, numberstyle=\color{mGray}, label={lst:query1}, numbers=left, basicstyle=\footnotesize, xleftmargin=2em,frame=single,framexleftmargin=1.5em]
WITH RECURSIVE edges_cte (id, from, to, column1, 
    ..., columnN, depth) AS 
    (SELECT edges.id, edges.from, edges.to, 
        edges.column1, ...,  edges.columnN, 0 
     FROM edges WHERE edges.id = 0
     UNION ALL
     SELECT edges.id, edges.from, edges.to, 
        edges.column1, ..., edges.columnN,
     e.depth + 1 FROM edges JOIN  edges_cte AS e
     ON edges.from = e.to AND e.depth < DEPTH_VAL)
SELECT edges_cte.id, edges_cte.from, edges_cte.to, 
    edges_cte.column1, ..., edges_cte.columnN
FROM edges_cte;
\end{lstlisting}
    
    \item For the third experiemnt, we have created a special type of query: 

    \begin{lstlisting}[language=sql, numberstyle=\color{mGray}, label={lst:query1}, numbers=left, basicstyle=\footnotesize, xleftmargin=2em,frame=single,framexleftmargin=1.5em]
WITH RECURSIVE edges_cte(id, to, depth) AS
    (SELECT edges.id, edges.to, 0 FROM edges
    WHERE edges.from = 0
    UNION ALL
    SELECT edges.id, edges.to, e.depth + 1
    FROM edges JOIN edges_cte AS e 
    ON edges.from = e.to AND e.depth < DEPTH_VAL)
SELECT edges.id, edges.to, edges.from,
    column1, ..., columnN FROM edges JOIN
    edges_cte ON edges.id = edges_cte.id;
    \end{lstlisting}
    
\end{enumerate}

In all plans of all evaluated systems the \texttt{edges\_cte} was hashed in the hash join (default PostgreSQL behavior).

\subsection{Experimental Setup}

Experiments were performed using the following hardware and software configuration. Hardware: Intel® Core™ i7-8550U CPU @ 1.80GHz (8 cores), 16 GiB RAM, 500GB Samsung PSSD T7. Software: Ubuntu 20.04.5 LTS x86\_64, Kernel 5.15.0-60-generic, gcc 9.4.0, PostgreSQL 14.2.

\subsection{Experiments \& Discussion}

\begin{figure*}[h!]
     \centering
     \includegraphics[width=0.99\linewidth]{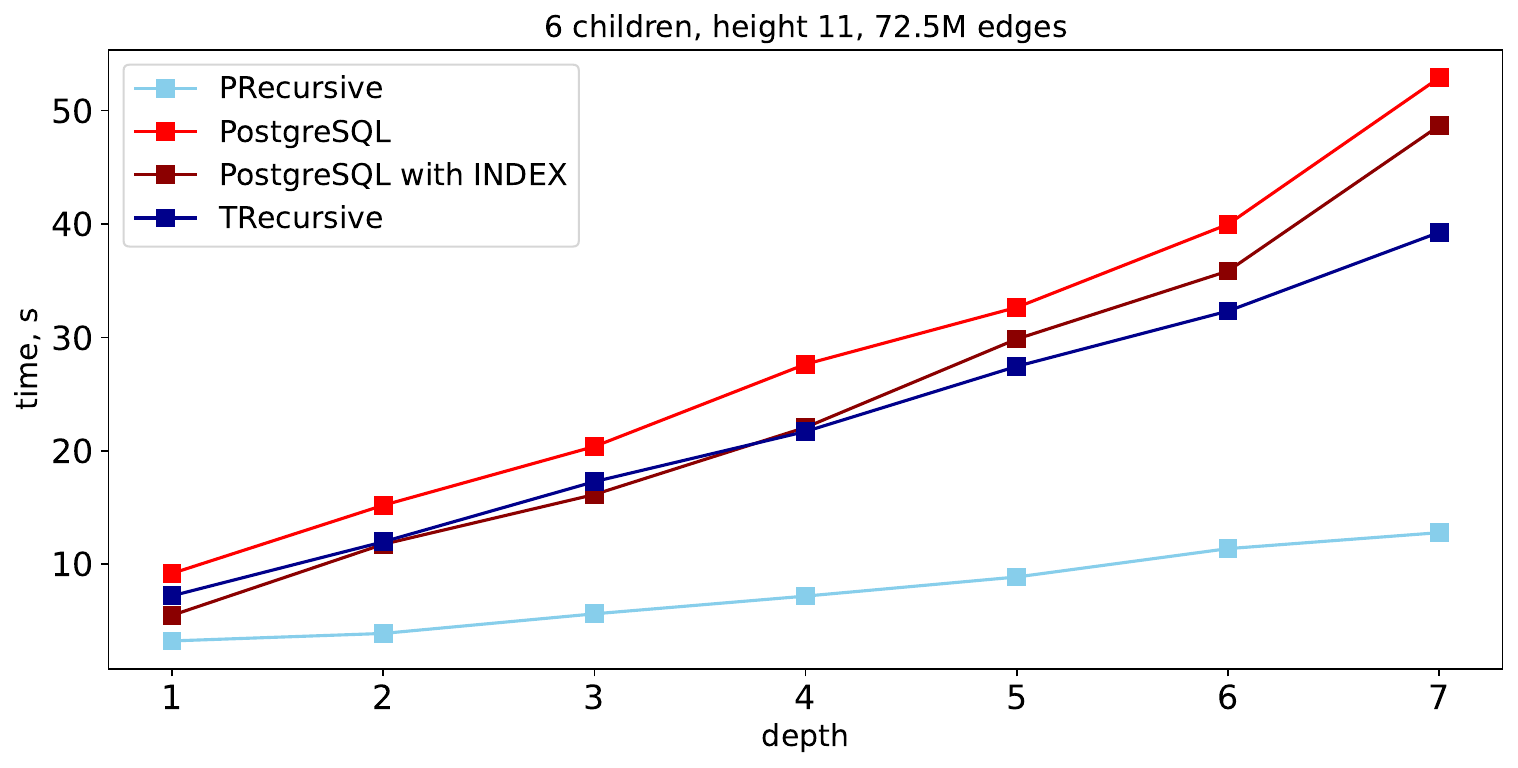}
     \caption{Experiment 1 results}
     \label{fig:exp1}
\end{figure*}

\textbf{Experiment 1.} The results are presented in Figure~\ref{fig:exp1}. The \texttt{TRecursive} approach exhibits performance that is similar to that of PostgreSQL, as expected, since the underlying query engines perform identical operations. Meanwhile, \texttt{PRecursive} outperforms \texttt{TRecursive} significantly, because it uses only two out of the four attributes (\texttt{from} and \texttt{to}) during the search, and materializes values of the third attribute (\texttt{id}) only when the required table rows are known. The number of rows scheduled to be materialized is much smaller than the total number of rows in the table (by roughly 200 times), resulting in operators passing around intermediate results of much smaller size. The index in PostgreSQL was built over \texttt{from}, since it is used to find edges in the join in the recursive part of the query. As we can see, this helps improve PostgreSQL performance, although it is a small improvement. Moreover, with the increase in depth, TRecursive demonstrates slightly better results compared to PostgreSQL with Index.

\textbf{Experiment 2.} The queries considered in this experiment were executed on a dataset with an additional parameter, denoted as \texttt{N}, which corresponds to the number of additional columns. The query plans for \PosDBAnonymized{} and PostgreSQL remain almost identical to those used in the first experiment, with the addition of ancillary columns to the \texttt{Materialize} operators in \PosDBAnonymized{} and to the projections in PostgreSQL, respectively.

\begin{figure*}[h!]
     \centering
     \includegraphics[width=0.99\linewidth]{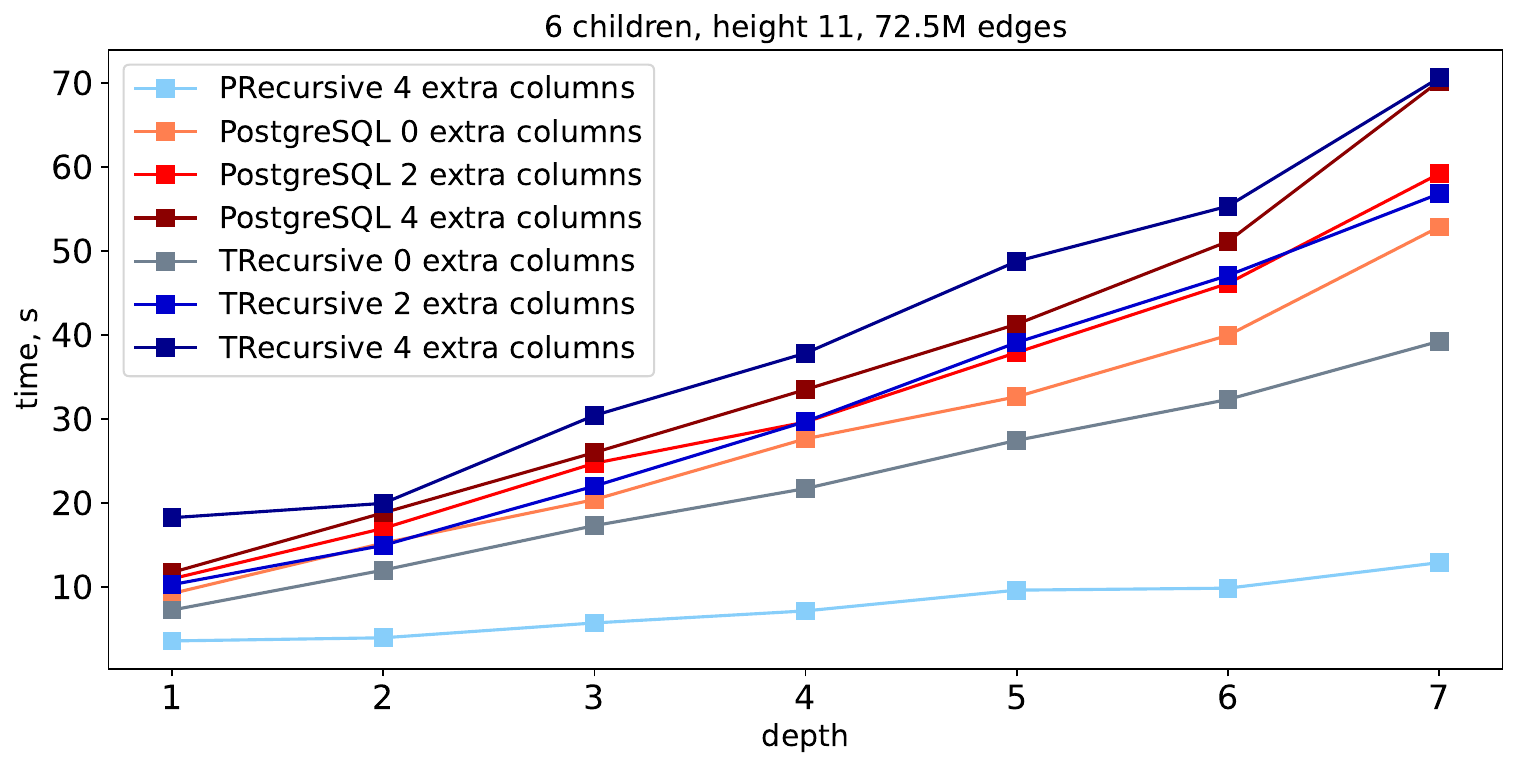}
     \caption{Experiment 2 results}
     \label{fig:exp2a}
\end{figure*}

Due to space constraints, we did not include \texttt{PostgreSQL with INDEX} in this and subsequent experiments, as its behavior is equivalent to that of PostgreSQL when changing the parameter \texttt{N}. Furthermore, results of \texttt{PRecursive} were only included for the maximum number of additional columns, as the time taken by \texttt{PRecursive} was predictably found to be almost independent of \texttt{N}.

The run times of the queries depending on the depth of the traversal are presented in Fig.~\ref{fig:exp2a}. As we can see, with increasing depth, the gap in the runtime on tables of different ``widths'' grows. \PosDBAnonymized{} \texttt{PRecursive} outperforms all other approaches.  This happens due to late materialization reducing sizes of intermediate results significantly. In this experiment, it matters even more due to the substantial overhead associated with passing ``wide'' intermediates (all columns) between operators, even though only two of them are required for the recursive part (\texttt{from} and \texttt{to}). It's important to mention that as the ``width'' of passed row grows, \PosDBAnonymized{} \texttt{TRecursive} falls behind PostgreSQL. This happens because of columnar nature of \PosDBAnonymized{}. With \texttt{TRecursive} it requires more disk accesses (one for each column) to retrieve a single row from a table. In contrast, PostgreSQL can do this with a single access since all the data for table rows is stored together.

\textbf{Experiment 3.} This experiment is similar to the previous one, but now we are trying to conserve space in \texttt{edges\_cte} by reducing the size of the intermediates. We only store the data necessary for reconstructing the original table row and navigating through the tree. This query requires more RAM since a second copy of the edges table is required by the top-level join. Therefore, we had to reduce the dataset due to the memory constraints of the employed hardware.

\begin{figure*}[h!]
     \centering
     \includegraphics[width=0.99\linewidth]{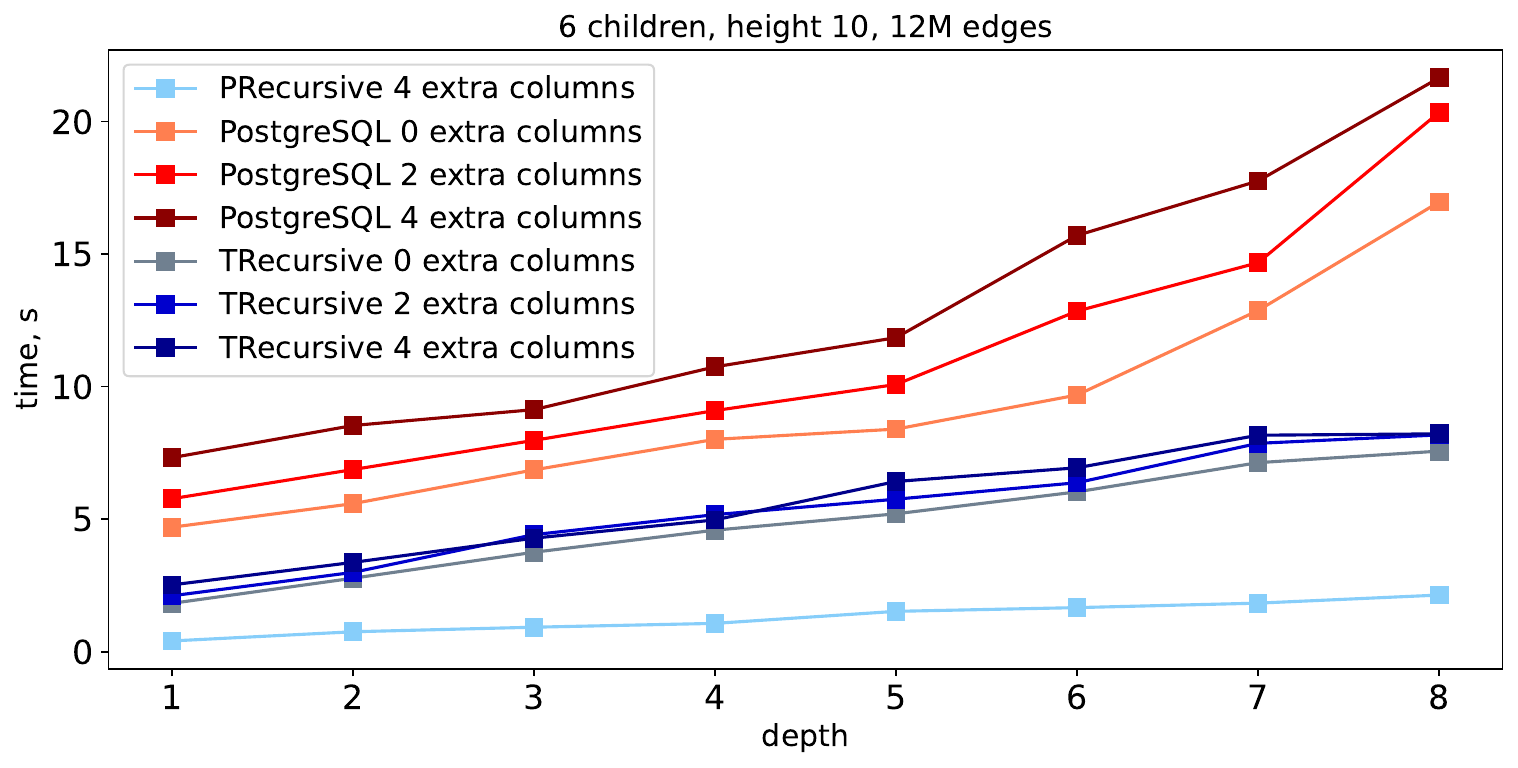}
     \caption{Experiment 3 results}
     \label{fig:exp2b}
\end{figure*}
The run times of the queries depending on the depth of the traversal are presented in Fig.~\ref{fig:exp2b}. It can be seen that the performance of \texttt{PRecursive} is similar to its performance in the previous experiment, it has the best performance in this experiment too among all compared approaches. \texttt{TRecursive}, however, beats PostgreSQL in this experiment.

As we can see by the performance improvement of \texttt{TRecursive}, this method helps reduce disk access overhead of row reconstruction inside the \texttt{TRecursive} operator. In \PosDBAnonymized{}, unnecessary columns are now only materialized once at the very end. Whereas in \texttt{PostgreSQL}, the internal hash join involves inefficient sequential data reads that discard unnecessary columns. Finally, this experiment shows that our approach cannot be emulated inside PostgreSQL via join.

\section{Conclusion}~\label{sec:concl}%
In this paper, we proposed two approaches to implementing recursive queries in a position-enabled column-oriented DBMS: \texttt{TRecursive} and \texttt{PRecursive}, with the latter utilizing positions to implement the late materialization approach. We implemented two sets of operators: 1) a tuple-based set which is similar to the operators that can be found in classical row-stores but leveraging columnar data access, and 2) a positional-based set which is the main contribution of the paper.

We conducted experiments to evaluate the performance of the proposed approaches and used PostgreSQL as the baseline. Experiments demonstrated that both approaches offer improvement, with \texttt{PRecursive} offering up to 6 times performance gain over PostgreSQL and 3 times over \texttt{TRecursive}. However, \texttt{TRecursive} remains the only option if there are two (or more) distinct tables in the \texttt{RECURSIVE} part used, due to implementation-related restrictions. Also, \texttt{TRecursive} yields up to 3x improvement over PostgreSQL when an additional payload columns which are not required in the \texttt{RECURSIVE} part exist and the query can be rewritten to project them only in the top-level. Finally, we shown that it is not possible to emulate our approach inside a row-store efficiently.

\bibliographystyle{splncs04}
\bibliography{bibliography-short}
\end{document}